%% file: eprint.tex
\documentclass[10pt]{article}
\usepackage{graphicx}
\usepackage{subfigure}
\usepackage{lineno}

\def\Title#1{\begin{center} {\Large #1 } \end{center}}
\def\Author#1{\begin{center}{ \sc #1} \end{center}}
\def\Address#1{\begin{center}{ \it #1} \end{center}}

\newcommand\pubblock{\rightline{\begin{tabular}{l} Proceedings of the Second Annual LHCP\\ \pubnumber\\
         \pubdate  \end{tabular}}}

\newenvironment{Abstract}{\begin{quotation} \begin{center} 
             \large ABSTRACT \end{center}\bigskip 
      \begin{center}\begin{large}}{\end{large}\end{center} \end{quotation}}

\newenvironment{Presented}{\begin{quotation} \begin{center} 
             PRESENTED AT\end{center}\bigskip 
      \begin{center}\begin{large}}{\end{large}\end{center} \end{quotation}}

\input econfmacros.tex

\usepackage{color}

\newcommand\pubnumber{ ATL-PHYS-PROC-2014-127 }

\newcommand\pubdate{\today}

\def\affiliation{
On behalf of the ATLAS Collaboration, \\
Department of Physics \\
Birmingham University, West Midlands B15 2TT, United Kingdom }

\begin{document}
\large
\begin{titlepage}
\pubblock

\vfill
\Title{   Combined measurements of the properties of the Higgs boson using the ATLAS detector  }
\vfill

\Author{ Ludovica Aperio Bella  }
\Address{\affiliation}
\vfill
\begin{Abstract}

The combined measurement  of the mass, couplings and spin-CP properties of the recently discovered Higgs boson obtained with the ATLAS detector using up to $25$ fb$^{-1}$ of $7$ TeV and $8$ TeV $pp$ collision data is reviewed.

\end{Abstract}
\vfill

\begin{Presented}
The Second Annual Conference\\
 on Large Hadron Collider Physics \\
Columbia University, New York, U.S.A \\ 
June 2-7, 2014
\end{Presented}
\vfill
\end{titlepage}
\def\thefootnote{\fnsymbol{footnote}}
\setcounter{footnote}{0}
%

\normalsize 


\section{Introduction}

The discovery of a new particle  with a mass of about 125 GeV, consistent with the SM Higgs boson,
 was announced by the ATLAS and CMS
Collaborations on July $4^{th}$ 2012  \cite{Aad:2012tfa},\cite{Chatrchyan:2012ufa}. Following its discovery, the properties of this
particle are studied using $4.5$ fb$^{-1}$ of $pp$ collision data at $\sqrt s = 7$ TeV and $20.3$ fb$^{-1}$ at $\sqrt s = 8$ TeV.
 Here, updated measurements of the mass and rates
of the observed new particle are presented using the ATLAS detector~\cite{ATLAS}. The measured yields are analysed in terms of the signal rates, for different production and decay
modes and for their combination. Finally, the couplings of the newly discovered boson are probed
with fits to the observed data under specific assumptions.

\section{Mass measurement}
\begin{figure}[!h]
\centering 
   \includegraphics[width=0.55\textwidth]{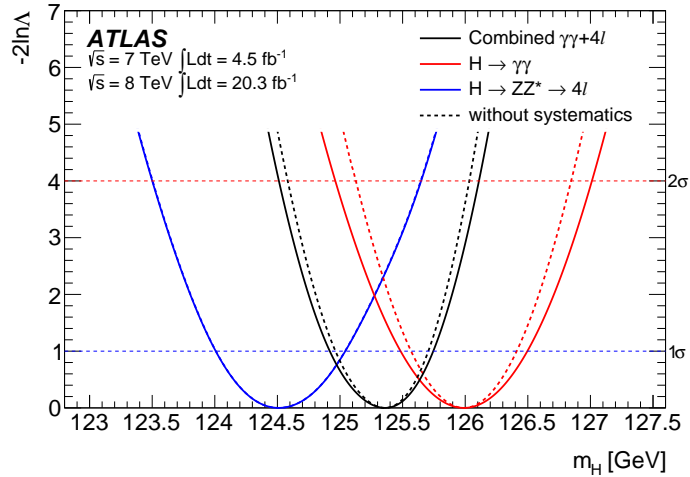}
   \caption{ {\small Value of $-2 \ln \Lambda$  as a function of $m_H$ for the $H\rightarrow \gamma\gamma$ and $H\rightarrow ZZ^{(*)} \rightarrow 4\ell$ channels individually and their combination, where the signal strengths $\mu_{\gamma\gamma}$ and $\mu_{4\ell}$ are allowed varying independently. The dashed lines show the statistical component of the mass measurements. For the $H\rightarrow ZZ^{(*)} \rightarrow 4\ell$ channel, this is indistinguishable from the solid line which includes the systematic uncertainties. 
 ~\cite{massPaper} } }
\label{fig:figurea}
\end{figure}

A new mass measurement of the Higgs boson is derived from a combined fit to the invariant mass spectra of the decay channels $H\rightarrow \gamma\gamma$ and $H\rightarrow ZZ^{(*)} \rightarrow 4\ell$.
This  result  supersedes the previous result from ATLAS~\cite{pl}.
The new $H\rightarrow \gamma\gamma$ result profits from an improved calibration
of the energy measurements of electron and photon candidates~\cite{calibeg}, which results in a substantial reduction of the systematic
uncertainties on their energy scales. In the $H\rightarrow ZZ^{(*)} \rightarrow 4\ell$ channel, both the statistical and the systematic uncertainties on the mass measurement have been reduced. 
The improvement on the statistical uncertainty arises primarily from the use of a multivariate discriminant, designed
to separate the signal from the continuum background. The systematic uncertainty reduction comes from both the
improved electromagnetic energy calibration and a reduction in the muon momentum scale uncertainty, which was
obtained by studying large samples of $Z\rightarrow \mu^+ \mu^-$ and $J/\psi \rightarrow \mu^+ \mu^-$ decays~\cite{calibmu}.
\\The mass spectra and the discriminating variables for the two channels are fitted simultaneously using an unbinned maximum likelihood fit with background and signal parameterisation described in Ref~\cite{massPaper}.
\\The measured values of the Higgs boson mass obtained in the two channels separately and for their combination is reported in Table \ref{tab:table1}. In Figure~\ref{fig:figurea}, the likelihood ($\Lambda$) scan as a function of $m_H$ for the individual channels and their combination are shown.
  \\
To quantify the consistency between the two measured Higgs mass values, a likelihood function
is considered for the mass difference $\Delta m_H = m^{\gamma\gamma}_H - m^{4\ell}_H $ with the averaged mass profiled in the
fit. The best fit value for the mass difference is  $\Delta m_{H} = 1.47 \pm 0.67 (stat) \pm 0.28 (syst)$ GeV, giving a compatibility of 4.8\% ($1.98 \sigma $) with the single resonance hypothesis.

\begin{table}[!h]
\begin{center}
\begin{tabular}{l|c}  
Channel &  Channel Mass measurement [GeV] \\ \hline
 $H\rightarrow \gamma\gamma$ & $125.98 \pm 0.42 (stat) \pm 0.28 (syst) = 125.98 \pm 0.50$ \\ 
$H\rightarrow ZZ^{(*)} \rightarrow 4\ell$ &  $124.51 \pm 0.52 (stat) \pm 0.06 (syst) = 124.51 \pm 0.52$\\ \hline
Combined & $125.36 \pm 0.37 (stat) \pm 0.18 (syst) = 125.36 \pm 0.41$\\ \hline
 \end{tabular}
\caption{ Summary of Higgs boson mass measurements. ~\cite{massPaper} }
\label{tab:table1}
\end{center}
\end{table}

\section{Signal strength measurements}

\begin{figure}[!h]
\centering 
    \subfigure[\label{fig:mu}]{\includegraphics[width=0.45\textwidth]{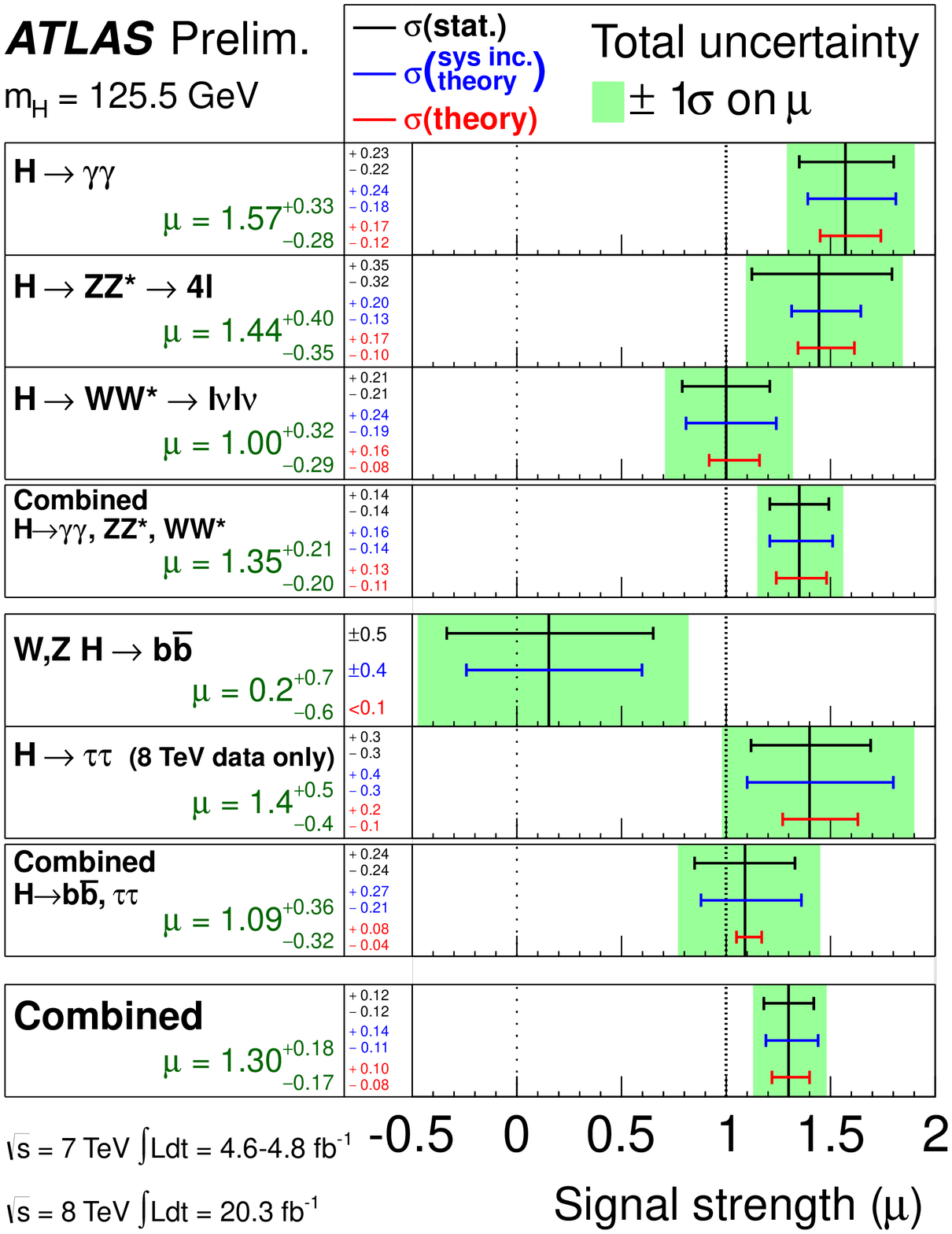}}
    \subfigure[\label{fig:cup}]{\includegraphics[width=0.45\textwidth]{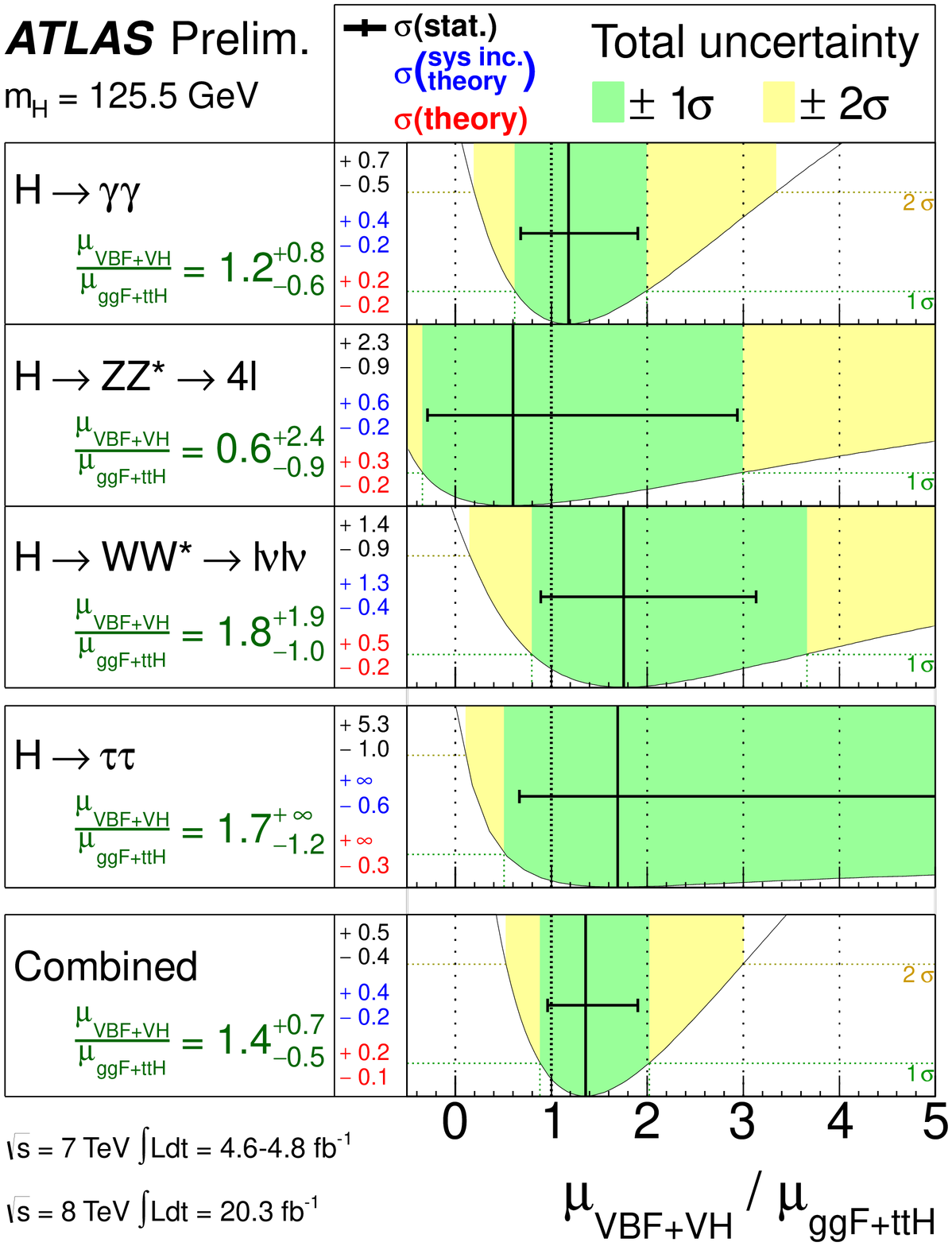}}
\caption{ {\small \subref{fig:mu} The measured signal strengths for a Higgs boson of mass $m_H =125.5$ GeV, normalised to the SM expectations, for the individual final states and various combinations. The best-fit values are shown by the solid vertical lines. The total $\pm1\sigma$ uncertainties are indicated by green shaded bands. 
\subref{fig:cup} Measurements of the $\mu_{VBF+VH}/\mu_{ggF+ttH}$ ratios for the individual final states and their combination, for a Higgs boson mass $m_H~=~125.5$~GeV. The best-fit values are represented by the solid vertical lines, with the total $\pm1\sigma$ and $\pm2\sigma$ uncertainties indicated by the green and yellow shaded bands, respectively, and the statistical uncertainties by the superimposed horizontal error bars.~\cite{couplingCONF}
} }
\label{fig:figure1}
\end{figure}

The measured yields from the Higgs boson decay channels have been analysed in terms of
signal strength, $\mu$, defined as the measured yields normalised to the SM prediction, for the different production and decay modes~\cite{couplingCONF}. Hypothesis testing and confidence
intervals are based on the profile likelihood ratio test statistic $\Lambda(\alpha)$, where $\alpha$ represents one or
more parameter of interest such as the Higgs boson signal strength. 
 This test statistic
extracts the information on the parameters of interest from the full likelihood function, assuming
a fixed common $m_{H}$ hypothesis corresponding to the measured value $m_H = 125.5$ GeV\footnote{The signal  strength results presented here have been obtained assuming as $m_{H}$ hypothesis the ATLAS mass measurement reported in Ref.~\cite{pl}.} .
The best-fit values of the signal strength parameter for each channel
independently and for the combination are shown in Figure~\ref{fig:mu}.
The combination of the two recently published $H\rightarrow b \bar{b}$ and $H\rightarrow \tau\tau$ results~\cite{fermion,fermion2}, yields a signal strength of
$\mu_{b\bar{b},\tau\tau} = 1.09 \pm 0.24(stat) ^{+0.27}_{-0.21}(syst)$, corresponding to $3.7 \sigma$ evidence for the direct decay of the Higgs boson into fermions. 
The global signal strength combining all five channels was found to be $\mu = 1.30 \pm 0.12(stat) ^{+0.14}_{-0.11}(syst)$. 
The compatibility with the SM is about 7\%.
A significant component of the systematic uncertainty of this measurement is associated to the theory expectation of the cross sections and branching ratios for the dominant production mode, gluon-gluon fusion process (ggF). The theory uncertainty is dominated by uncertainties on the QCD renormalisation and factorisation scales and the parton distribution function (PDF). \\
The measurements of the signal strengths described above do not give direct information on the relative contributions of the different production mechanisms. 
Therefore, in addition to the signal strengths of different decay channels, the signal strengths of different production processes contributing to the same decay channel are determined, exploiting the sensitivity offered by the use of event categories in the analyses of all the channels.
The data are fitted separating the production processes involving the Higgs boson coupling to vector bosons (vector boson fusion (VBF) and associated production (VH)) from the gluon mediated production modes (ggF and ttH), involving the Higgs boson coupling to fermions to gain direct information on the relative contributions of the different production mechanisms. 
A model independent test of the theory can be performed by 
taking the ratio of the two signal strengths $\mu_{VBF+VH}/\mu_{ggF+ttH}$ for the individual final states and their combination as shown in Figure~\ref{fig:cup}.
A value of $\mu_{VBF+VH}/\mu_{ggF+ttH} = 1.4^{+0.5}_{-0.4}(stat)^{+0.4}_{-0.2}(syst)$ is obtained from the combination of the $H\rightarrow \gamma\gamma$, $H\rightarrow ZZ^{(*)} \rightarrow 4\ell$, $H\rightarrow W W^{(*)}$ and $H\rightarrow \tau\tau$  channels. 
To test the sensitivity to VBF alone the ratio $\mu_{VBF}/\mu_{ggF+ttH}$ is measured independently.
 The fit result is  $\mu_{VBF}/\mu_{ggF+ttH} = 1.4^{+0.5}_{-0.4}(stat)^{+0.4}_{-0.3}(syst)$ providing an evidence at the $4.1\sigma$ level that a fraction of Higgs boson production occurs through VBF.

\section{Coupling studies}

For a consistent treatment of Higgs boson couplings in production and decay modes, coupling scale factors are defined as multiplicative modifiers 
following the approach and benchmarks recommended in Ref.~\cite{Xsec}. 
The signal is assumed to originate from a single narrow resonance of mass $m_H = 125.5$ GeV\footnote{All the coupling results presented here have been obtained assuming as $m_{H}$ hypothesis the ATLAS mass measurement reported in Ref.~\cite{pl}.} 
with a negligible width and only modifications of couplings strengths are taken into account, while the tensor structure of the couplings is assumed to be the same as in the SM. 
In this way for all Higgs channels, the products $\sigma \times BR(i\rightarrow H\rightarrow f)$ can be decomposed as $\sigma \times BR(i\rightarrow H\rightarrow f) = \frac{\sigma_{i} \Gamma_f}{\Gamma_H} $, where ${\Gamma_f}$ is the partial decay width into the final state $f$ and ${\Gamma_H}$ is the total width of the Higgs.
Following this prescription, the LO-motivated coupling scale factors $\kappa_j$ are defined in such a way that the cross section $\sigma_j$ and the partial decay width ${\Gamma_j}$ associated with the SM particle $j$ scales as $\kappa^2_j$ with respect to the corresponding SM prediction. \\
The results are then extracted from fits to the data using the profile likelihood ratio $\Lambda(\kappa)$, where the couplings scale factors  $\kappa_j$ are treated either as parameters of interest or as nuisance parameters, depending on the scenario~\cite{couplingCONF}.

\subsection{Couplings to fermions and bosons}
\begin{figure}[!h]
\centering 
   \includegraphics[width=0.55\textwidth]{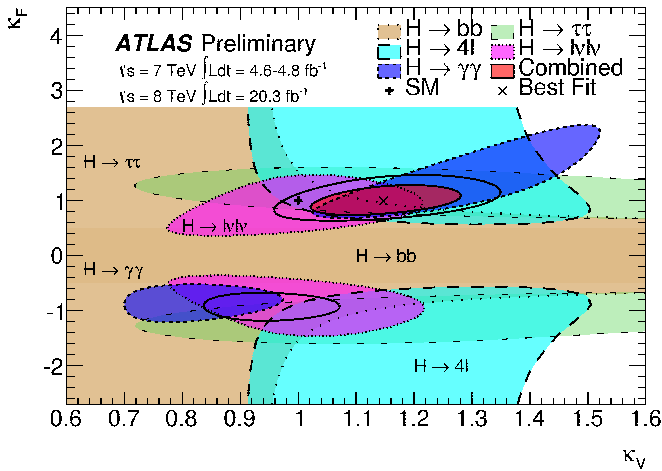}
\caption{ {\small Correlation of the coupling scale factors $\kappa_F$ and $\kappa_V$ overlaying the $68\%$ CL contours derived from the individual channels and their combination. 
~\cite{couplingCONF} }}
\label{fig:k}
\end{figure}

The first benchmark model considered here assumes a common coupling scale factor for fermions, $\kappa_F$ , and one for bosons, $\kappa_V$.
The fit is performed in two variants, with and without the assumption that the total width of the Higgs boson is given by the sum of the known SM Higgs boson decay modes.
For the first benchmark model, the 2D scan in the ($\kappa_V$, $\kappa_F$) plane is shown in Figure~\ref{fig:k} with the $68\%$ CL contours derived from the individual channels and their combination. The best-fit values are $\kappa_V = 1.15\pm0.08$ and $\kappa_F = 0.99^{+0.17}_{-0.15}$, and the two-dimensional compatibility with the SM hypothesis is $12\%$. \\
In the second benchmark, the fit is repeated without assumption on the total Higgs width. In this case only ratios of coupling scale factors can be measured, therefore the free parameters are defined as $\lambda_{FV} = \frac{\kappa_F}{\kappa_V }$ and $\kappa_{VV} = \kappa_V \cdot \kappa_V /\kappa_H$, where $\lambda_{FV} $ is the ratio of the fermion and vector boson coupling scale factors, and $\kappa_{VV}$ is an overall scale that includes the total width and applies to all rates. The measured values are $\lambda_{FV} = 0.86^{+0.14}_{-0.12}$ and $\kappa_{VV} = 1.28^{+0.16}_{-0.15}$, and the two-dimensional compatibility of the SM hypothesis with the best-fit point is $10\%$.

\subsection{Custodial Symmetry and relations within the fermion coupling sector}
Identical coupling scale factors for the $W$ and $Z$ boson are required within tight bounds by the $SU(2)$ custodial symmetry. To test this constraint directly in the Higgs sector, the ratio $\lambda_{WZ} = \kappa_W/\kappa_Z$ is probed. 
The best fit result obtained considering only SM contributions, is found to be $\lambda_{WZ} = 0.94^{+0.14}_{-0.29}$. This result is compatible with the SM hypothesis within $19\%$.

The currently accessible fermionic Higgs decay channels allow probing also the relations between the up- and down-type quarks. First, the ratio $\lambda_{du}$ between
down- and up-type fermions is probed, while vector boson couplings are taken unified. Around the SM-like minimum, $\lambda_{du} = 0.95^{+0.20}_{-0.18}$, this fit provides a $\sim3.6\sigma$ level evidence of the coupling of the Higgs boson to down-type fermions.
Then the ratio $\lambda_{lq}$ between leptons and quarks coupling modifiers has also been probed. Around the SM like
minimum, $\lambda_{lq}  = 1.22^{+0.28}_{-0.24} $. With this result, a vanishing coupling of the Higgs boson to leptons is excluded at the $\sim4\sigma$  level.

\subsection{Probing beyond the SM contributions}
%

In all the studies previously presented, the couplings of the SM particles to the Higgs boson are assumed to be as predicted by the SM. 
To measure possible extra contributions from new particles in Higgs coupling vertices including loops, $H\rightarrow \gamma\gamma$ and $gg \rightarrow H$, effective scale factors $\kappa_\gamma$ and $\kappa_g$ are introduced. The potential new particles contributing to the $H\rightarrow \gamma\gamma$ and $gg \rightarrow H$ loops may contribute to the total width of the Higgs. 
The resulting variation in the total width is parameterised in terms of the additional branching ratio into invisible or undetected particles ($BR_{i.,u.}$) as the following: $\Gamma_H = \frac{\kappa^2_H(\kappa_i)}{(1-BR_{i.u.})} \Gamma^{SM}_H  $. 
Fitting the data and using the physical constraint $BR_{i.,u.} > 0$, the $95\%$ CL upper limit on the additional branching ratio into invisible or undetected particles is found to be $BR_{i.,u.} < 0.41$ (the SM expected limit is $BR_{i.,u.} < 0.55$). 

\begin{figure}[!ht]
\centering 
   \includegraphics[width=0.5\textwidth]{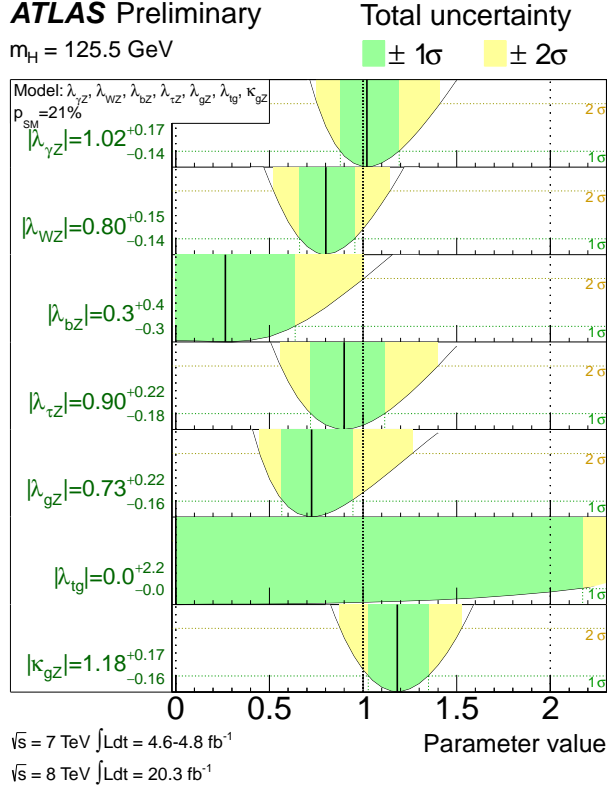}
   \caption{ {\small Summary of the coupling scale factor measurements in the generic models with independent $\kappa_\gamma$, $\kappa_g$ and no assumption on the total width. The best-fit values are represented by the solid vertical lines, with the total $\pm1\sigma$ and $\pm2\sigma$  uncertainties indicated by the green and yellow shaded bands, respectively. For each model, the n-dimensional compatibility of the SM hypothesis with the best-fit point is given by $p_{SM}$.
 ~\cite{couplingCONF}} }
\label{fig:figure3}
\end{figure}

\subsection{Generic models}
Finally, generic coupling fits without assumptions on the relationships between the coupling scale factors, 
allow potential deviations from the SM hypothesis to be searched for. In the previous benchmark models, specific aspects of the Higgs sector were tested by combining coupling scale factors into a minimum number
of parameters sensitive to the probed scenario. 
Within the following generic models, the couplings scale factors to $W, Z, t, b$ and $\tau$ are treated independently, while for the $H\rightarrow \gamma\gamma$ decay and $gg \rightarrow H$ production and the total width $\Gamma_H$, either the SM particle content is assumed or no assumptions are made.
Therefore, the coupling scale factor fit in this more general model uses only few basic assumptions and hence represents the most model-independent determination of coupling scale factors that is currently possible.  All results of this fit, summarised in Figure~\ref{fig:figure3}, are in good agreement with the SM expectation.

\subsection{Constraints on New Phenomena via Higgs Boson Coupling Measurements}
The observed rates in the different channels are also used, for example, to determine the mass dependence of the Higgs boson couplings to other particles. 
The coupling scale factors to fermions and vector bosons are expressed in terms of a mass scaling factor $\epsilon$ and a \emph{vacuum expectation value} parameter $M$~\cite{BSM_mass}.
The best-fit point was found to be compatible with the SM expectation and the measured couplings to fermions and vector bosons are consistent with the linear and quadratic mass dependence predicted in the SM. 

Furthermore, the rate measurements are used to derive direct limits on beyond-Standard-Model (BSM) theories~\cite{BSMcouplingCONF}. 
Here one example is given considering a simplified Minimal Supersymmetric Standard Model (MSSM)~\cite{susy1,susy2,susy3}.
In this example, the Higgs coupling scale factors to fermions and vector bosons are expressed as a function of two parameters of interest: $m_A$ and $\tan\beta$ .
The two-dimensional likelihood scan in the $(m_A ,\tan\beta)$ plane for this simplified MSSM model shows that the data are consistent with the SM hypothesis.   
 The observed (expected) lower limit at $95\%$ CL on the CP-odd Higgs boson mass was found to be $m_A > 400$ GeV ($280$ GeV) for $2 \leq \tan \beta \leq 10$, with the limit increasing to larger masses for $\tan \beta < 2$.

\section{Conclusions}
A review of the latest results on the main properties of the Higgs boson has been presented. 
An improved measurement of the mass of the Higgs boson was derived from a combined fit to the invariant mass spectra of the decay channels $H\rightarrow \gamma\gamma$ and $H\rightarrow ZZ^{(*)} \rightarrow 4\ell$. The measured value of the Higgs boson mass is $m_H = 125.36 \pm 0.37 (stat) \pm 0.18 (syst)$ GeV. This result is based on improved energy-scale calibrations for photons, electrons, and muons as well as other analysis improvements, and supersedes the previous result from ATLAS.\\
The combination
of the major Higgs decay channels: $H\rightarrow \gamma\gamma$, $H\rightarrow ZZ^{(*)} \rightarrow 4\ell$, $H\rightarrow W W^{(*)}$, $H\rightarrow \tau\tau$ and  $H\rightarrow b \bar{b}$ results in a signal strength value of
$\mu = 1.30 \pm 0.12(stat) ^{+0.14}_{-0.11}(syst)$. Moreover, evidence for Higgs decaying to fermions is found at the $3.7\sigma$ level using the newly available channels  $H\rightarrow b \bar{b}$ and $H\rightarrow \tau\tau$.
The compatibility of the measured yields for the studied channels with the predictions for the SM
Higgs boson is tested under various benchmark assumptions probing salient features of the couplings.

\end{document}

%% file: econfmacros.tex
\def\beq{\begin{equation}}
\def\eeq#1{\label{#1}\end{equation}}
\def\eeqn{\end{equation}}

\def\beqa{\begin{eqnarray}}
\def\eeqa#1{\label{#1}\end{eqnarray}}
\def\eeqan{\end{eqnarray}}

\let\bar=\overbar

\def\Dslash{\not{\hbox{\kern-4pt $D$}}}
\def\dslash{\not{\hbox{\kern-2pt $\del$}}}

\def\msb{{\bar{\ssstyle M \kern -1pt S}}}

